# Experience gravity in the classroom using the rubber sheet: an educational proposal from the collaboration between University and School

Adriana Postiglione[1,2], Ilaria De Angelis[1,2]

[1] Dipartimento di Matematica e Fisica, Università Roma Tre, Rome (ITALY)
[2] INFN Sezione di Roma Tre, Rome, (ITALY)

E-mail: adriana.postiglione@uniroma3.it

**Abstract**

Teaching modern physics in high school is of increasingly importance as it can offer students a more realistic and updated vision of the world, and can provide an opportunity to understand the most recent scientific discoveries. In this context, General Relativity (GR) occupies a prominent place, since it is related to astonishing scientific results, such as the first image of a black hole or the discovery of gravitational waves. In this paper we describe an educational proposal aimed at teaching GR in high school in a fun and playful way using the so-called rubber sheet analogy. We present a set of instructions to build a simple and low-cost space-time simulator, and a series of related educational cards that guide the teacher in the implementation of the activities step by step. This work is the result of a long and productive debate among Italian high school teachers who collaborate since many years with the Department of Mathematics and Physics of Roma Tre University in Rome. As our proposal relies on the collaboration with the final users, we believe that it will meet their needs and expectations, and it will help to treat GR in high school more and more over time.

## 1. Introduction

In recent years, an increasing attention is given to the teaching of modern physics in high school [1-8]. The reason is twofold: on one hand, students should embrace a more realistic and updated vision of the world; on the other hand, they should also be put in the position of understanding, at least at a qualitative level, the latest scientific discoveries reported in the newspaper front pages. Introducing modern physics in high school requires that the teachers are able to handle the topics with sufficient confidence and autonomy, and to promote activities that intrigue their students and make them understand very complex aspects. This is not an easy task, for which the support of the educational research carried on by Universities can be beneficial.

In this contest, a special place is occupied by General Relativity (GR), the theory introduced by Albert Einstein in 1916 [9], from which astonishing discoveries have been made, including the first image of a black hole [10] and the discovery of gravitational waves [11]. In order to introduce GR in schools and investigate students' understanding, several approaches have been explored: online learning environments [12], geometrical approaches [13], simplified mathematical treatments [14], thought experiments [15], investigating the understanding of younger students [16, 17], focusing on gravitational waves [18, 19], warped time [20, 21], geodesics motion [22].

However, in most cases these proposals did not rely on an initial co-planning of the activity between university researchers and teachers. Conversely, the present work proposes an educational path that arises from a long and profitable interaction with high school teachers, aimed at meeting their needs and expectations and making them autonomous in bringing the activity into their classes.

The model used in this work is the popular rubber sheet analogy (RSA), which seems to have been introduced by Einstein himself [23] in the attempt of better explaining his theory. According to GR, gravity comes out from the deformation of the four dimensional space-time, which is not a static stage, but rather a dynamical element that can change depending on the objects placed in it: a massive object, such as a star, deforms space-time producing gravity, i.e. forcing the planets to orbit around it. In the RSA, the fabric of space-time is compared to a stretched rubber sheet; if a small weight is placed on the sheet, it produces a warp in it; if someone throws a marble, this latter would fall towards the weight, eventually turning around it before falling.

The RSA has been massively adopted to talk about GR in a simple way [24-27]. It is known that it can be misleading for some aspects, since it hides that space-time is four-dimensional, that it has a temporal dimension, that the curvature around massive objects is symmetric in all dimensions and that in the Universe there is no friction like there is on the sheet [28-31]. However, it remains an "excellent analogy" [32], a powerful activity able to attract young people [18] and definitely a teaching tool whose potential we believe is still worth to be explored.

For this reason, within the activities we are carrying on with schools at the Department of Mathematics and Physics of Roma Tre University about GR, we have extensively used the RSA. Our goal is to build an educational path suitable for high schools, highlighting which aspects of GR could be explained using different weights and marbles and to what extent.

We started building a space-time simulator using a circular structure of 1.8 m of diameter that could support a lycra sheet, and we tested it with different targets. Then, we started to involve teachers to have some feedback and to better understand their needs in order to replicate the activity in complete autonomy. After months of tests, debates, doubts and proposed solutions, we managed to package an educational path approved by teachers that allows to explore gravity through the RSA, and that teachers can carry out and extend over different high school grades, alternating practical activities with examples and insights. We identified a simple and low-cost kit that allows to build a rubber sheet space-time suitable to be used in a classroom, and we built a series of educational cards that guide the user step by step, suggesting the right weights and marbles to use, the best ways to throw them, the most effective way to approach the topics.

In order to make all this material easier to consult, we have collected it in an open access eBook [33] that could be easily downloaded by a large audience of teachers and educators.

The rest of the paper is organized as follows. In section 2 we describe the educational and outreach contexts where we used the RSA, and the development of our DO-IT-YOURSELF (DIY) kit. In section 3 we describe the educational cards we produced and the instruction to build the space-time with low-cost materials. In section 4 we discuss our work, while in section 5 we sum up our results and show some future prospects.

**2. Project background and methods**

Firstly we tested the concept, designing and building a first version of the space-time setup, in order to explore potentials and weaknesses of the educational experiment. The setup was made at the Department of Mathematics and Physics of Roma Tre University, in collaboration with the mechanical shop of the INFN Roma Tre Section and consisted of an aluminium circular structure having a 1.8 meters diameter and covered by a 2x1.5 meters lycra sheet. To keep the sheet tightly stretched we used some aluminium hooks.

The RSA model was tested for the first time during one of the public outreach events organized by the Department in 2017. This type of events is particularly suitable to test new educational displays, since it typically hosts a large number of persons of all ages, who discover several science topics in a playful and fun atmosphere, through several simultaneous activities. During that event children, teenagers and adults were all intrigued by the experiment and willingly participated in throwing marbles on the sheet. We also notice that, although the average time of participation at other stands was about 15 minutes, some of the participants stayed at the space-time display much longer, or they came back several times during the public outreach event. The audience reaction to the new exhibit was thus very positive, and thanks to the playful approach used, they were willing to listen and ask questions about complex topics related to gravity, such as black holes or gravitational waves. Given such positive feedback, we decided to adopt the same playful and interactive approach also when the exhibit was organized in a high school classroom.

In the second phase, we focused on a more structured activity suitable for schools that could have been used to approach different topics such as Kepler's laws, gravity assist, gravitational lensing and black holes. More than 200 high school students of different ages participated to the educational activity; the groups consisted of students belonging to the same class (often the entire class) or to students coming from different classes but selected by their teachers because they were very interested in these scientific topics. We administered pre- and post- questionnaires to the participants, in order to explore their reactions and comprehension of the different topics according to their age. The results of such tests, which will be described in detail in a future paper [34], show a remarkable improvement in the knowledge of the topics addressed. The ones usually treated in school, such as Kepler's laws, are more easily visualized, better understood and remembered longer; the advanced topics, such as black holes and gravitational lenses, are well understood by almost all the participants after the activity and remembered also after months.

At the beginning of 2019 we started a systematic collaboration with a group of teachers who were attending the annual course on Modern Physics for high school teachers offered by the Department. During the course, we presented the activity and trained the 25 participants to the use the exhibit. The teachers were very interested in learning how to use the RSA for teaching GR, but they also highlighted the importance of detailed guidelines that could support them, step by step, during the performance of the experiment in the classroom. Moreover, they pointed out that the space in the classrooms are usually small and the resources are limited, so they need a space-time simulator that can fit in a classroom, it is inexpensive and easy to build. After several weeks of discussions, suggestions, and design improvements, also involving other high school teachers in the project, at the beginning of 2020 we produced a low-cost DIY kit that we consider simple, inexpensive, easy-to-build but still didactically valid. We also developed a series of related educational cards that could guide the teacher in carrying out the activity. Since these cards were produced on the basis of the needs of the final user, they contain all the tricks and guidelines necessary to properly select and use the marbles and weights on the rubber sheet, in relation to the physical phenomena that should be described during the lesson.

**3. Results**

Overall we developed eight educational cards that guide teachers step by step in carrying out different activities focused on phenomena related to gravity, using the rubber sheet. The topics covered by the cards are gradually more complex: from the Newtonian gravity, that is typically already treated in high schools, to topics that can be fully explained only with GR. In order to be easily usable by teachers, these cards have been collected in the open access eBook[1] "*Sperimentare la gravità con il telo elastico: linee guida e trucchi. Experience gravity with the rubber sheet: guidelines and tricks.*" [33].

The book has been written both in Italian, in order to facilitate its use by Italian teachers, and in English, in order to ensure a wide accessibility to all possible users.

The cards deal with activities created to be suitable for high school students, especially for the ones who are already familiar with Kepler's laws (which in the Italian school system typically occurs during the third year, that corresponds to students of 15-16 years old).

Each card contains the topic addressed, a brief description of the purpose of the proposed activity, a list of the materials necessary to realize the activity, and a list of the steps necessary to show the phenomenon. All the tricks and guidelines to choose the right marbles and weights and to properly launch them on the sheet are specified. For each card there is also a section that collects notes and insights, in which images, videos and links are recommended, to carry out a more detailed activity. Tips and advices are specified in order to make students' understanding more effective.

*3.1 Description of the educational cards*

In this section we list the topics addressed in each of the eight educational cards we developed.

The first two cards provide a general introduction to the exhibit, describing the structure to be built and the model used. The following three ones focus on phenomena that can be explained by the Newtonian theory of gravity, i.e. without introducing the concept of space-time; however, the RSA helps students to more correctly visualize and experience the phenomena. The remaining three cards treat topics that only GR can correctly explain, which also represent the most recent achievements of research in astrophysics, such as gravitational lensing and black holes. In this case a significant use of supporting materials, such as properly chosen photos and videos, is made in order to supplement the activity.

The first card describes the minimum kit required to build the rubber sheet space-time. The basic version we propose requires a hula-hoop of at least 80 cm in diameter, a 1x1 meter lycra sheet, six wooden slats that act as support for the structure, some strews, clothespins, tube clips, and a selection of marbles and weights.

The second card provides an introduction to the space-time as described by the rubber sheet analogy; a particular attention is dedicated to its weaknesses [28-31], so that the teacher is fully aware of the model he is using.

The third card is dedicated to the way planets orbit around the Sun, and allows to treat Kepler's laws. In the card some tricks are suggested to the user in order to better show the peculiarities expressed in each law.

The fourth card focuses on the so-called gravity assist, the phenomenon for which an object accelerates when approaching a massive body, like a planet, due to the gravitational attraction. Using the lycra sheet, it is indeed possible to show that the marble accelerates when approaching the central weight. This gives the opportunity to talk about the way a space probe travels inside the Solar System.

---

[1] The eBook can be downloaded at the following link: http://www.edizioniefesto.it/collane/circuli-dimensio/379-sperimentare-la-gravita-con-il-telo-elastico-linee-guida-e-trucchi-experience-gravity-with-the-rubber-sheet-guidelines-and-tricks

The fifth card of the book deals with binary systems, i.e. the systems consisting of two stars orbiting around each other; placing two weights on the lycra sheet and throwing some marbles, it is possible to make them orbit around the weights like a planet belonging to a binary system orbits around the two stars.

With the sixth card, teachers can guide their students to the discovery of the phenomenon of gravitational lensing and its incredible consequences. The marble thrown on the lycra sheet, in fact, deflects its trajectory when approaching the central weight. Similarly, a ray of light is bended when it travels near a massive object like a star or a galaxy. From this simple observation, teachers can start a debate on what an observer would see if a massive object were located between him and the source of light. Then, through the videos and photos mentioned in the card, further insights can be made.

The seventh card focuses on black holes, probably the most famous and fascinating objects of modern astrophysics. The first part of the activity deals with the visualization on the rubber sheet of the compactness, one of the main features of black holes, together with the fact that more or less compact objects, even with the same mass, deform the space-time in a different way. Then, the simulation of a black hole is done through a small but very heavy weight (like a small magnets pair tied to a weight placed under the sheet), so that the teacher can show that the marbles rotating very close to it reach very high speeds, higher than the ones reached around a less compact object. Finally, recalling what has been learned with the sixth card, it is possible to explain the first image of a black hole published in 2019 [10] or why the black hole *Gargantua* looks like it is in the film Interstellar [35].

The eighth and last educational card of the book suggests additional ideas about the use of the lycra sheet. It describes how to simulate the formation of the Solar System, how to give an idea of the time dilation and how to visualize gravitational waves.

The complete cards can be found at the link: http://www.edizioniefesto.it/images/openaccess/Postiglione-De-Angelis---Sperimentare-la-gravita-con-il-telo-elastico.pdf

*3.2 Evaluations of future readers' response*

On July 2020 the book was published. We are going to track the response to the book through two questionnaires that readers are invited to answer at the end. The first questionnaire concerns an initial appraisal of the eBook: if the description of the activities is clear and exhaustive, the materials provided are sufficient, and the reader plans to replicate the activity in the classroom. The second questionnaire is instead dedicated to those teachers who carried out the activities and can thus evaluate the feasibility of realization, the students' reaction and the efficacy in explaining the physics concepts.

## 4. Discussion

Although gravity is massively treated in high schools, the approach used is old, since we know that the concepts introduced by Newtonian theory do not correctly describe the reality. Moreover, in order to talk about the recent discoveries of modern physics, the introduction of GR is mandatory.

A complete and rigorous treatment of GR in high schools is implausible due to the advanced mathematical knowledge required, but it still can be proposed focusing on its key concept, that John Wheeler beautifully synthesized in the sentence: "*mass tells spacetime how to curve, and spacetime tells mass how to move*". The model of the RSA allows to visualize this key feature, thus helping comprehend the space-time without resorting to equations.

Our eBook follows our strong conviction of the importance of introducing GR into school through the RSA.

In fact, although the usage of the rubber sheet to simulate the space-time shows some weaknesses (addressed with the second educational card of our eBook), thanks to the debate with the teachers we came to the conclusion that it can actually be useful to visualize and experiment phenomena described both by the Newtonian theory of gravity and its modern counterpart.

We believe that it is precisely this collaboration with teachers that represents the most interesting aspect of our work, since it ensured that our proposals were in line with the needs of the final users. For the same reason, we believe that the material we developed can help to insert in a fairly structured way the treatment of GR in high schools, increasing the number of teachers who could use the RSA in their classrooms.

A first confirmation of all this comes from the fact that, as soon as the course finished in February 2020 (before Italian schools were closed due to Covid-19 lockdown), three of the teachers who attended it immediately built the DIY structure, thus long before the publication of the book. Two of them followed step by step the model proposed in our first educational card (shown in Figure 1), while one of them built an improved version that is also closable (see Figure 2).

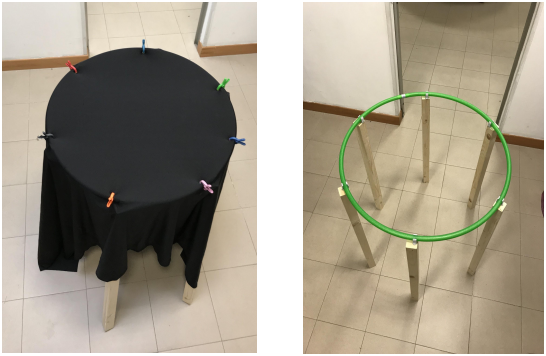

Figure 1. The model of the space-time structure we propose in the first educational card of the book ("Sperimentare la gravità con il telo elastico: linee guida e trucchi. Experience gravity with the rubber sheet: guidelines and tricks" [33]). The materials used to build the structure are cheap and easy to find.

## 5. Summary and conclusion

In this paper we presented an educational path that aims to help high school teachers to deal with GR in a practical and playful way through a series of educational cards and a DIY kit that exploit the rubber sheet analogy, and that have been collected in an open-access eBook [33].

The first feedback we received from teachers led us to think that it is actually possible to talk about GR in high schools, and that our eBook can be a valid ally to do that.

Since the book includes questionnaires that invite the reader to give an evaluation of the educational proposal, it also represents a tool that would enrich and further improve the proposed activities, also providing further clues on the feasibility of dealing with GR in high schools.

In the next future we will continue to test our proposal in two ways: on the one hand, we will closely follow the work of the teachers who closely collaborated with the Department and are willing to realize these activities in their classrooms;

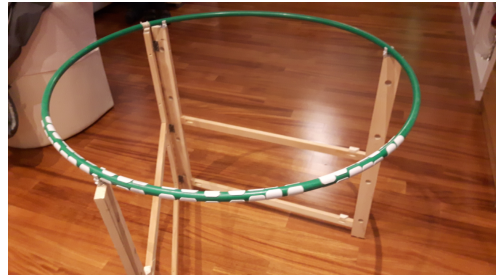

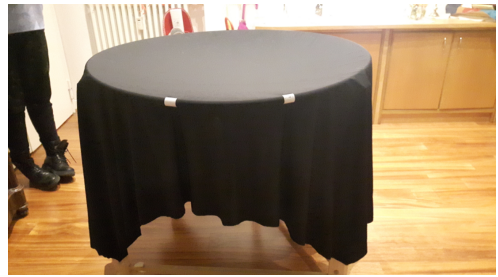

Figure 2. An improved and closable version of the space-time rubber sheet as created by one of the teachers after seeing our proposal.

on the other hand, we will monitor the evaluations of the activities as provided by the readers of the book, which had already been downloaded more than 200 times in early October 2020.

Finally, we also plan to create tutorial videos that can accompany teachers in building the space-time structure and carrying out the activities.

## Acknowledgements

This work came to life thanks to the support of the Italian Project "Piano Lauree Scientifiche" and the Young Minds Section of Rome of the European Physical Society. The authors would like to thank the mechanical shop of the INFN Roma Tre Section for building our first wonderful space-time structure. Also thanks to the teachers who participated to the high school teacher refresher course held at the Department of Mathematics and Physics of Roma Tre University, for their incredible willingness to share their experiences and knowledge. Thanks to professor Dan Burns of the "Los Gatos" high school in Los Angeles for having shared his enlightening video[2] on YouTube.

---

[2] www.youtube.com/watch?v=MTY1Kje0yLg